# How to strictly calculate self-consistent fields of realistic plasma particles

H. Lin

*State Key Laboratory of High Field Laser Physics, Shanghai Institute of*

*Optics and Fine Mechanics,*

*P. O. Box 800-211, Shanghai 201800, China;*

*linhai@siom.ac.cn*

()

Because currently the most popular method of calculating plasma self-consistent fields is an incomplete method which, strictly speaking, is not suitable to scientific investigation, we develop this method into be a complete reliable basic tool for scientific investigation.

Plasma physics is a physical branch about many charged particles interacting through their self-consistent fields. In its earlier developing stage (about 1940s~1960s), many theoretical methods which are successful in other elder physical branch such as neutral gas physics and fluid mechanics were transplanted into this younger branch and rapidly built up the basis of this new branch. For example, in almost all basic textbooks in plasma physics [1-6], plasma wave is studied by not only non-relativistic fluid dynamics but also various analytical ansatzs on microscopic Vlasov-Maxwell equations. However, almost no one doubts whether these transplanted methods are appropriate for plasmas where numerous charged particles are correlated through their self-consistent fields. More important, in above-mentioned transplanted methods the plasma self-consistent fields is never strictly calculated but is indeed treated by various (obvious and hidden) approximations.

Following example illustrates clearly a typical inconsistency of above-mentioned analytical ansatzs on microscopic Vlasov-Maxwell equations. In many basic textbooks, people make Fourier analysis: $f_1 = \sum_k f_k \exp(i\theta); E_1 = \sum_k E_k \exp(i\theta);$ and $\theta = kr - \omega t$ on Vlasov

equation+Maxwell equations (in $B = 0$ case)

$$0 = \partial_t f_1 + v \cdot \nabla f_1 - E_1 \cdot \partial_p f_0; \tag{1.a}$$

$$\nabla \cdot E_1 = \int f_1 d^3 p \tag{1.b}$$

$$\partial_t E_1 = \int v f_1 d^3 v. \tag{1.c}$$

Especially, when $E_1$ is of a monochromic wave form $E_1 = Cons * \exp[i(kr - \omega t)]$, this will lead to well-known Landau damping which is an important conception for plasma wave [1-6]. On the other hand, if $E_1 = Cons * \exp[i(kr - \omega t)]$, Eq.(1.b,c) will lead to

$$\int v f_1 d^3 v = \frac{\omega}{k} \int f_1 d^3 v$$

and the fluid velocity meeting $u = \frac{\int v f d^3 v}{\int f d^3 v} = \frac{\omega}{k} * \frac{\int f_1 d^3 v}{\int f_1 d^3 v + \int f_0 d^3 v}$. However, after deriving fluid momentum equation from Vlasov equation according to standard procedure, one could find following relation

$$\partial_t \left[ \frac{\omega}{k} * \frac{\int f_1 d^3 v}{\int f_1 d^3 v + \int f_0 d^3 v} \right] * \int f_1 d^3 v + \text{convective. term} = E_1 * \int f_0 d^3 v. \tag{2}$$

This suggests that if the phase velocity $\frac{\omega}{k}$ is a constant, the shape of the wave, or the shape of $n = \int f_1 d^3 v$, have to meet two equations, the continuity equation and Eq.(2) at $\partial_t \frac{\omega}{k} = 0$ case, and an assumed shape might not meet these two equations, which are both able to be expressed completely in term of $n = \int f_1 d^3 v$. Therefore, a reliable investigation on a plasma wave should be based on strictly calculating self-consistent fields rather than some empirical ansatzs, which might be inconsistent to studied model equations.

Although strictly calculating self-consistent fields represents a correct direction in plasma physics, the detailed technical road to achieve this goal causes this correct goal being greatly discounted. The chosen detailed technical road is the well-known particle simulation scheme, in which particles' information and fields' information are alternatively updated through numerous Newton equations and Maxwell equations [7-9]. This inevitably encounters a realistic question: the number of realistic particles is an astronomical figure and hence corresponds to too huge data mount. It is impossible to update so huge mount of data even one time.

Therefore, this inevitably causes a merging approximation, which means $N$ realistic particles being merged into a so-called macroparticle and hence cut down significantly corresponding data mount, being introduced into the particle simulation scheme. Unfortunately, although this merging approximation makes updating being feasible, it will also cause a hidden disaster which refers to calculated $E$ and $B$ are functions of the merging ratio $R_{merge} = N : 1$. People of course wish that with $N$ decreasing, the dynamics of calculated macroparticles is more and more approaching to the dynamics of real particles. Therefore, according to this viewpoint, it is reasonable to take the dynamics of calculated macroparticles of sufficiently small size as the dynamics of realistic particles. Unfortunately, this viewpoint is not true. It is almost impossible to give a mathematically strict proof on the uniform convergence of $E(r, t; R_{merge} = N : 1)$ to $E(r, t; R_{merge} = 1 : 1)$ with $R_{merge}$ decreasing. Let us see an equation set including $2N + 4$ equations describing $N$ realistic particles (or macroparticles)

$$\partial_t E(R,t) = \nabla \times B(R,t) + \sum_i d_t r_i(t) \delta(r_i(t) - R); \tag{3.1}$$

$$\partial_t B(R,t) = -\nabla \times E(R,t); \tag{3.2}$$

$$\nabla \cdot E(R,t) = \sum_i \delta(r_i(t) - R); \tag{3.3}$$

$$\nabla \cdot B(R,t) = 0; \tag{3.4}$$

...

$$d_t \frac{d_t r_i(t)}{\sqrt{1 - [d_t r_i(t)]^2}} = E(r_i(t), t) + d_t r_i(t) \times B(r_i(t), t) \tag{3.2i+4}$$

$$v_i = d_t r_i \tag{3.2i+5}$$

...

If we merge two real particles into a macroparticle (or two macroparticles into a macro-macroparticle), we indeed deal with another equation set also including $2N + 4$ equations

$$\partial_t E(R,t) = \nabla \times B(R,t) + \sum_i d_t r_i(t) \delta(r_i(t) - R); \tag{4.1}$$

$$\partial_t B(R,t) = -\nabla \times E(R,t); \tag{4.2}$$

$$\nabla \cdot E(R,t) = \sum_i \delta(r_i(t) - R); \tag{4.3}$$

$$\nabla \cdot B(R,t) = 0; \tag{4.4}$$

...

$$d_t \frac{d_t r_i(t)}{\sqrt{1-[d_t r_i(t)]^2}} = E(r_i(t),t) + d_t r_i(t) \times B(r_i(t),t) \tag{4.4k+4}$$

$$v_{2k} = d_t r_{2k} \tag{4.4k+5}$$

$$d_t \frac{d_t r_{2k+1}(t)}{\sqrt{1-[d_t r_{2k+1}(t)]^2}} = d_t \frac{d_t r_{2k}(t)}{\sqrt{1-[d_t r_{2k}(t)]^2}} \tag{4.4k+6}$$

$$v_{2k+1} = d_t r_{2k+1} \tag{4.4k+7}$$

...

Here, if $i$ is even or $= 2k$, Eq.(3.2i+4) and Eq.(3.2i+5) are just Eq.(4.4k+4) and Eq.(4.4k+5). But if $i$ is odd or $= 2k+1$, Eq.(3.2i+4) and Eq.(3.2i+5) will be replaced by Eq.(4.4k+6) and Eq.(4.4k+7). Obviously, the difference between these two equation sets is significant (because nearly half of total equations are approximated) and their respective solutions also have marked difference. More important, no matter how large $N$ is, the difference between above two equation sets, and that between their solutions, is still significant. Therefore, it is impractical to expect following relation being valid

$$d_N \left\{ \sup_{r,t} |E(r,t;R_{merge}=N:1) - E(r,t;R_{merge}=1:1)| \right\} > 0 \tag{5}$$

This relation, which implies that the smaller $N$ is, the closer $E(r,t;R_{merge}=N:1)$ is to $E(r,t;R_{merge}=1:1)$, is just the theoretical basis of above-mentioned viewpoint.

However, it is easy to obtain $E(r,t;R_{merge}=1:1)$ if we note a fact: Any solution of following equation set of $2N+5$ members

$$0 = \left[ d_t \frac{u(r_i(t),t)}{\sqrt{1-[u(r_i(t),t)]^2}} - E(r_i(t),t) - u(r_i(t),t) \times B(r_i(t),t) \right] + \Pi \tag{6.0}$$

$$\partial_t E(R,t) = \nabla \times B(R,t) + \sum_i d_t r_i(t) \delta(r_i(t) - R); \tag{6.1}$$

$$\partial_t B(R,t) = -\nabla \times E(R,t); \tag{6.2}$$

$$\nabla \cdot E(R,t) = \sum_i \delta(r_i(t) - R); \tag{6.3}$$

$$\nabla \cdot B(R,t) = 0; \tag{6.4}$$

...

$$\left[ d_t \frac{d_t r_i(t)}{\sqrt{1-[d_t r_i(t)]^2}} - d_t \frac{u(r_i(t),t)}{\sqrt{1-[u(r_i(t),t)]^2}} \right] - \Pi = [d_t r_i(t) - u(r_i(t),t)] \times B(r_i(t),t) \tag{6.2i+4}$$

$$v_i = d_t r_i \tag{6.2i+5}$$

...;

where $u(R,t) = \sum_{i \in r_i(t)=R} d_t r_i(t) / \sum_{i \in r_i(t)=R} 1$, must also be a solution of Eqs.(3). In a mathematical language, Eqs.(3) and Eqs(6) have their respective solution sets: {solutions of Eqs.(6)} and {solutions of Eqs.(3)}, and there strictly exists a relation between these two sets: {solutions of Eqs.(6)} $\subset$ {solutions of Eqs.(3)}. On the other hand, one can easily find, by subtracting any two Newton equations in Eqs.(3), following relation for any two particles

$$d_t \frac{d_t r_i(t)}{\sqrt{1-[d_t r_i(t)]^2}} - d_t \frac{d_t r_j(t)}{\sqrt{1-[d_t r_j(t)]^2}}$$
$$= E(r_i(t),t) + d_t r_i(t) \times B(r_i(t),t) - E(r_j(t),t) + d_t r_j(t) \times B(r_j(t),t). \tag{7}$$

If two particles are at a same space position and the velocity of one particle is just equal to the fluid velocity at this position, Eq.(7) will automatically return to Eq.(6.2i+4) or Eq.(6.2j+4) at $\Pi = 0$ case. This fact implies that any solution of Eqs.(3) also meets Eqs.(6) at $\Pi = 0$ case, i.e., {solutions of Eqs.(3)} $\subset$ {solutions of Eqs.(6) at $\Pi = 0$ case}. Because of the relation {solutions of Eqs.(6) at $\Pi = 0$ case} $\subset$ {solutions of Eqs.(6)} $\subset$ {solutions of Eqs.(3)}, thus, we have {solutions of Eqs.(3)} = {solutions of Eqs.(6) at $\Pi = 0$ case}.

Eqs.(6) at $\Pi = 0$ case, a closed fluid equation set of $u$, $E$ and $B$, implies a fast and exact method of calculating $E(r,t;R_{merge}=1:1)$. Indeed, if people had noted following strict relation several decades ago, the chosen detailed technical road will be free from above-mentioned disaster. Let us see well-known equation for fluid momentum $p_{fl} = \int p f d^3 p / \int f d^3 p$ [1-6]

$$\partial_t p_{fl} + u_{fl} \cdot \nabla p_{fl} = E + u_{fl} \times B + \text{thermal pressure/density} \tag{A.1}$$

where $u_{fl}$ is the fluid velocity $u_{fl} = \int v f d^3 p / \int f d^3 p$. It is a pity for the community of plasma physics during several decades to fail of finding a very simple short-cut from above equation. Because the velocity $v$ is a nonlinear function of the momentum $p$ (i.e., $v = \frac{p}{\sqrt{1+p^2}}$) and vice versa, we should note that the statistic average value $\int p f d^3 p / \int f d^3 p$ (i.e. fluid momentum) is usually not equal to the momentum corresponded by the statistic average value $\int v f d^3 p / \int f d^3 p$ (or fluid velocity), i.e., $p_{fl} \neq p(u_{fl})$ (where $p(u_{fl})$ refers to the value of function $p(variable)$ at $variable = u_{fl}$), if the distribution $f$ is not a Dirac function of $p$ (i.e., $f$ has a thermal spread over $p$-space). Only at zero temperature case, there is $p_{fl} = p(u_{fl})$. (Strictly speaking, if $f$ is a symmetric function of $p$, there will be $p_{fl} = p(u_{fl}) = 0$, $u_{fl} = 0$ and thermal pressure$\neq 0$. But this special case corresponds to $E = 0$ and $E + u_{fl} \times B = 0$. A non-zero thermal pressure will drive $p_{fl}$ differing from 0 according to Eq.(A.1). Once $p_{fl} \neq 0$, there will be $p_{fl} \neq p(u_{fl})$ because $f$ has an asymmetric thermal spread over $p$-space).

Any distribution function $f$ has two independent characteristic parameters: the variance and the mean. Here, the mean of $f$ is represented by $u_{fl}$. As above discussed, $p_{fl}$ is thus a binary function :$p_{fl} = p_{fl}(u_{fl}, variance)$. Thus, we could express $p_{fl}$ as a series of $vari$: $p_{fl} = \sum_{i \geqslant 0} \partial^i_{vari} p_{fl}|_{vari=0} (vari)^i$. Moreover, a term in Eq.(A.1), thermal pressure/density, could also be expressed as a similar series of $vari$ but this series does not contain $(vari)^0$-term because when $vari = 0$, $f$ is a Dirac function and hence thermal pressure is equal to zero. Namely, thermal pressure/density$= \sum_{i \geqslant 1} c_i (vari)^i$. Note that Eq.(A.1) is valid for any value of $vari$. Then by substituting these two series into Eq.(A.1) and comparing coefficients of different order $(vari)^i$-terms, we could find an equation set of infinite members and every member is for a definite order $(vari)^i$-term (because the fact that Eq.(A.1) is valid for any value of $vari$ requires these coefficients of different $(vari)^i$-terms being zero separately). For zero-order term ($(vari)^0$-term), we have $0 = \partial_t[p_{fl}(u_{fl},0)] + u_{fl} \cdot \nabla_r[p_{fl}(u_{fl},0)] - [E + u_{fl} \times B]$. Indeed, all equations for non-zero-order terms ($(vari)^{i \neq 0}$-terms) could be merged into an equation if we substracting Eq.(A.1) and the equation for zero-order term ($(vari)^0$-term), and this equation reads $0 = \partial_t[p_{fl} - p_{fl}(u_{fl},0)] + u_{fl} \cdot \nabla_r[p_{fl} - p_{fl}(u_{fl},0)]$+thermal pressure/density. Note that $p_{fl}(u_{fl},0) = p(u_{fl}) = \frac{u_{fl}}{\sqrt{1-u_{fl}^2}}$. Namely, because Eq.(A.1) is of a general form $0 = function(u_{fl}, vari)$, the conditions for such a general form being valid at arbitrary *vari*-value are: $0 = function(u_{fl}, 0)$ and $d_{vari} function = 0$. For Eq.(A.1), these two conditions could be expressed by two equations

$$0 = \partial_t[p_{fl} - p_{fl}(u_{fl}, 0)] + u_{fl} \cdot \nabla_r[p_{fl} - p_{fl}(u_{fl}, 0)] + \text{thermal pressure/density}; \quad \text{(A.2.a)}$$

$$0 = \partial_t[p_{fl}(u_{fl}, 0)] + u_{fl} \cdot \nabla_r[p_{fl}(u_{fl}, 0)] - [E + u_{fl} \times B], \quad \text{(A.2.b)}$$

and the former equation depends on the temperature (or *vari*) whereas the latter is independent of the temperature. Obviously, the solution of Eqs.(A.2) must be the solution of Eq.(A.1). Thus, $u_{fl}$, $E$ and $B$ form a closed description.

Obviously, the closed equation set (Eq.(A.2.b)+4 Meqs) has been misjudged as only valid at zero-temperature case for a long time, and therefore is often judged as an "approximation theory". People are easy to believe that the thermal pressure will inevitably affect $p_{fl}$ and *"further affect self-consistent fields (through Meqs)"*. Actually, a scrupulous reader will not have such a misunderstanding because he could notice that it is $u_{fl}$, rather than $p_{fl}$, that appears in *Meqs*. Moreover, Eq.(A.2.a) will lead to infinite correlated quantities which are all dependent on the variance of $f$. Like the treatment in many standard textbooks [1-6], introducing an assumed thermodynamics state equation could cut the chain of infinite correlated quantities and form a closed equation set. But this will again depreciate greatly the validity of fluid theory. Actually, fluid theory could become a nearly perfect basic tool in plasma physics according to following way: First, figuring out $E$ and $B$ of realistic particles and $u_{fl}$ through Eq.(A.2.b)+4 Meqs. Second, if one is interested in more detailed particles' information (such as particles trajectories and velocities, or so-called phase-space snapshots), he could solve Vlasov equation under known $E$ and $B$, or make a "test-particle calculation" on macroparticles, and then extract relevant information from solved $f$ or solved trajectories and velocities of macroparticles. More important, it is no need to introduce an assumed thermodynamics state equation. Eq.(A.2.a) itself actually implies a complicated thermodynamics state equation.

A more direct way of obtaining above closed equation set of $u_{fl}$, $E$ and $B$ could be start from Vlasov equation (VE). For any distribution $f$, we could expressed as $f = n\delta(v - u_{fl}) + a_0\delta(v - u_{fl}) + \sum_i a_{i\geqslant 1}(v - u_{fl})^i$ (where $n = \int f d^3p$, $u_{fl} = \int v f d^3p / \int f d^3p$, $a_i$ are independent of $v$, $a_0$ depends on all coefficients $a_{i\geqslant 1}$ through a relation $\int \left[ a_0\delta(v - u_{fl}) + \sum_{i\geqslant 1} a_i(v - u_{fl})^i \right] d^3p = 0$, i.e. $a_0$ is a function of all $a_{i\geqslant 1}$, $a_0 = a_0(a_1, ..., a_i, ...)$). Substituting this expression into VE and comparing the coefficients of $(v - u_{fl})^i$-term, we could find that there exists following equation for $f_{mono} = n\delta(v - u_{fl}) + a_0\delta(v - u_{fl})$ (because of the fact that VE is valid for any distribution $f$ which is characterized by its own coefficient set $\{a_i\}$)

$$\partial_t f_{mono} + u_{fl} \cdot \nabla f_{mono} - [E + u_{fl} \times B] \cdot \partial_p f_{mono} = 0, \tag{A.3}$$

which could directly lead to Eq.(A.2.b) according to standard procedure.

Moreover, there is an easier way of obtaining this closed equation set of $u_{fl}$, $E$ and $B$ from the starting equations of particle simulation. Note that a relativistic Newton equation

$$\begin{aligned}
0 &= d_t \frac{d_t r_i(t)}{\sqrt{1 - [d_t r_i(t)]^2}} - E(r_i(t), t) - d_t r_i(t) \times B(r_i(t), t) \\
&= \left[ \left[ d_t \frac{d_t r_i(t)}{\sqrt{1 - [d_t r_i(t)]^2}} - d_t \frac{u(r_i(t), t)}{\sqrt{1 - [u(r_i(t), t)]^2}} \right] - [d_t r_i(t) - u(r_i(t), t)] \times B(r_i(t), t) \right] \\
&+ \left[ d_t \frac{u(r_i(t), t)}{\sqrt{1 - [u(r_i(t), t)]^2}} - E(r_i(t), t) - u(r_i(t), t) \times B(r_i(t), t) \right],
\end{aligned} \tag{B.1}$$

is valid for arbitrary value of $d_t r_i(t)$, or arbitrary value of $\Delta = d_t r_i(t) - u(r_i(t), t)$. Because Eq.(B.1) is of a general form $0 = function(u(r_i(t), t), \Delta = d_t r_i(t) - u(r_i(t), t))$, the conditions for such a general form being valid at arbitrary $\Delta$-value are: $0 = function(u, 0)$ and $d_\Delta function = 0$. For Eq.(B.1), these two conditions could be expressed by two equations,

$$0 = \left[ \left[ d_t \frac{d_t r_i(t)}{\sqrt{1 - [d_t r_i(t)]^2}} - d_t \frac{u(r_i(t), t)}{\sqrt{1 - [u(r_i(t), t)]^2}} \right] - [d_t r_i(t) - u(r_i(t), t)] \times B(r_i(t), t) \right]; \tag{B.2.a}$$

$$0 = \left[ d_t \frac{u(r_i(t), t)}{\sqrt{1 - [u(r_i(t), t)]^2}} - E(r_i(t), t) - u(r_i(t), t) \times B(r_i(t), t) \right]. \tag{B.2.b}$$

where $u(R, t)$ represents the average value of the velocities of macroparticles whose positions are at $R$ when the time is $t$, i.e., $u(R,t) = \sum_{i \in r_i(t)=R} d_t r_i(t) / \sum_{i \in r_i(t)=R} 1$. Obviously, Eq.(B.2.b) is just equal to Eq.(A.2.b).

This also implies that even in the particles simulation scheme, it is no need to alternatively update particles' data and fields' data because fields' data could be strictly updated through Eqs.(6.0-5) at $\Pi = 0$ case. Here, alternative updating is the basic reason for the particles simulation scheme being entangled with above-mentioned disaster. This elder road characterized by alternatively updating, is indeed an approximation on the starting model equations Eqs.(3). If this disaster is not overcome, strictly speaking, particle simulation scheme is an incomplete method not suitable to scientific investigation. Here, this disaster could be easily overcome if another detailed road, which does not involve any approximation on Eqs.(3), is adopted, i.e., directly solving $E$ and $B$ from Eqs.(6.0-5) at $\Pi = 0$ case, rather than alternatively updating fields and particles. In short, no matter which one of the particles simulation scheme, the relativistic fluid theory, and the microscopic Vlasov-Maxwell theory is chosen by people as the starting model of investigating plasma physics, the self-consistent fields, $E$ and $B$, obey a fixed fluid equation set, i.e., Eqs(6.0-5) at $\Pi = 0$ case or Eq.(A.2.b)+4 Meqs. Indeed, these different basic methods are equivalent if they are in their respective strict forms. There is no reason to think that any method is better than others.

*A more concise and straightforward presentation of above text after the paragraph around Eq.(7) is given as below:*

Eqs.(6.0-4) at $\Pi = 0$ case, a closed fluid equation set of $u$, $E$ and $B$, implies a fast and exact method of calculating $E(r, t; R_{merge} = 1 : 1)$.

Indeed, if we notice the universality of some physical laws, we could obtain this closed equation set of $u$, $E$ and $B$ from different basic theoretical methods in plasma physics. At present, there are mainly three basic theoretical methods: fluid theory, Vlasov-Maxwell (VM) theory and above-mentioned particle simulation (or well-known PIC method). In every basic theoretical method, except 4 Maxwell equations (Meqs), other equations correspond to a basic physical law.

1.In fluid theory, this basic physical law is represented by fluid momentum equation [1-6]

$$0 = \partial_t p_{fl} + u_{fl} \cdot \nabla p_{fl} - E + u_{fl} \times B - \text{thermal pressure/density}$$
$$= [\partial_t (p_{fl} - p(u_{fl})) + u_{fl} \cdot \nabla (p_{fl} - p(u_{fl})) - \text{thermal pressure/density}]$$
$$+ [\partial_t p(u_{fl}) + u_{fl} \cdot \nabla p(u_{fl}) - E + u_{fl} \times B] ; \quad (A.1)$$

where $u_{fl}$ is the fluid velocity $u_{fl} = \int v f d^3 p / \int f d^3 p$ and $p_{fl} = \int p f d^3 p / \int f d^3 p$ is the fluid momentum. Because the velocity $v$ is a nonlinear function of the momentum $p$ (i.e., $v = \frac{p}{\sqrt{1+p^2}}$) and vice versa, we should note that the statistic average value $\int p f d^3 p / \int f d^3 p$ (i.e. fluid momentum) is usually not equal to the momentum corresponded by the statistic average value $\int v f d^3 p / \int f d^3 p$ (or fluid velocity), i.e., $p_{fl} \neq p(u_{fl})$(where $p(u_{fl})$ refers to the value of function $p(variable)$ at $variable = u_{fl}$), if the distribution $f$ is not a Dirac function of $p$ (i.e., $f$ has a thermal spread over $p$-space). Only at zero temperature case, there is $p_{fl} = p(u_{fl})$. (Strictly speaking, if $f$ is a symmetric function of $p$, there will be $p_{fl} = p(u_{fl}) = 0$, $u_{fl} = 0$ and thermal pressure$\neq 0$. But this special case corresponds to $E = 0$ and $E + u_{fl} \times B = 0$. A non-zero thermal pressure will drive $p_{fl}$ differing from 0 according to Eq.(A.1). Once $p_{fl} \neq 0$, there will be $p_{fl} \neq p(u_{fl})$ because $f$ has an asymmetric thermal spread over $p$-space).

2. In VM theory, this basic physical law is represented by Vlasov equation [1-6]

$$0 = \partial_t f + v \cdot \nabla f - [E + v \times B] \cdot \partial_p f.$$

$$= [\partial_t(f - f_{mono}) + v \cdot \nabla (f - f_{mono}) - [E + v \times B] \cdot \partial_p (f - f_{mono})]$$

$$+ (v - u_{fl}) \cdot \nabla f_{mono} - (v - u_{fl}) \cdot \partial_p f_{mono}$$

$$+ [\partial_t f_{mono} + u_{fl} \cdot \nabla f_{mono} - [E + u_{fl} \times B] \cdot \partial_p f_{mono}] \tag{B.1}$$

Any distribution function $f$ has two independent characteristic parameters: the variance and the mean. Here, the mean of $f$ is represented by $u_{fl} = \int v f d^3p / \int f d^3p)$. For any distribution $f$, we could expressed as $f = n\delta(v - u_{fl}) + a_0 \delta(v - u_{fl}) + \sum_{i \geqslant 1} a_i (v - u_{fl})^i$ (where $n = \int f d^3p$, $u_{fl} = \int v f d^3p / \int f d^3p$, $a_i$ are independent of $v$, $a_0$ depends on all coefficients $a_{i \geqslant 1}$ through a relation $\int \left[ a_0 \delta(v - u_{fl}) + \sum_{i \geqslant 1} a_i (v - u_{fl})^i \right] d^3p = 0$, i.e. $a_0$ is a function of all $a_{i \geqslant 1}$, $a_0 = a_0(a_1, ..., a_i, ...)$). The variance, $vari$, depends on the values of all $a_{i \geqslant 1}$.

3. In particle simulation, this basic physical law is represented by relativistic Newton equation of any particle [7-9]

$$0 = d_t \frac{d_t r_i(t)}{\sqrt{1 - [d_t r_i(t)]^2}} - E(r_i(t), t) - d_t r_i(t) \times B(r_i(t), t)$$

$$= \left[ \left[ d_t \frac{d_t r_i(t)}{\sqrt{1 - [d_t r_i(t)]^2}} - d_t \frac{u(r_i(t), t)}{\sqrt{1 - [u(r_i(t), t)]^2}} \right] - [d_t r_i(t) - u(r_i(t), t)] \times B(r_i(t), t) \right]$$

$$+ \left[ d_t \frac{u(r_i(t), t)}{\sqrt{1 - [u(r_i(t), t)]^2}} - E(r_i(t), t) - u(r_i(t), t) \times B(r_i(t), t) \right], \tag{C.1}$$

where $u(R, t)$ represents the average value of the velocities of macroparticles whose positions are at $R$ when the time is $t$, i.e., $u(R, t) = \sum_{i \in r_i(t) = R} d_t r_i(t) / \sum_{i \in r_i(t) = R} 1$.

The common of Eq.(A.1), Eq.(B.1) and Eq.(C.1) is that they have a binary-function type general form

$$0 = function(var1, var2) \tag{8}$$

where $var1$ and $var2$ are independent variables. 1. For fluid theory, because both $p_{fl} - p(u_{fl})$ and thermal pressure/density depend on $vari$ (i.e., the variance of $f$), there are $var1 = u_{fl}$ and $var2 = vari$ (i.e., the variance of $f$). Note that thermal pressure/density=

$\sum_{i\geqslant 1} c_i (vari)^i$ because when $vari = 0$, $f$ is a Dirac function and hence thermal pressure is equal to zero and hence this series does not contain $(vari)^0$-term. Likewise, $p_{fl}$ is thus a binary function : $p_{fl} = p_{fl}(u_{fl}, vari)$ and could be expressed as a series of $vari$: $p_{fl} = \sum_{i\geqslant 0} \partial^i_{vari} p_{fl}|_{vari=0} (vari)^i$ and $p_{fl}(u_{fl}, 0) = p(u_{fl}) = \frac{u_{fl}}{\sqrt{1-u_{fl}^2}}$. 2. For VM theory, there are $var1 = f_{mono}$ and $var2 = v - u_{fl}$. 3. For particle simulation, there are $var1 = u(r_i(t), t)$ and $var2 = d_t r_i(t) - u(r_i(t), t)$.

The universality of these basic physical laws requires that they are valid for arbitrary value of $var2$. For example, one cannot expect that a particle does not meet relativistic Newton equation when it corresponds to $var2 = d_t r_i(t) - u(r_i(t), t) = 0$ (because following equation is absurd)

$$\left[ d_t \frac{u(r_i(t), t)}{\sqrt{1 - [u(r_i(t), t)]^2}} - E(r_i(t), t) - u(r_i(t), t) \times B(r_i(t), t) \right] \neq 0.$$

Likewise, following equation is also absurd

$$[\partial_t p(u_{fl}) + u_{fl} \cdot \nabla p(u_{fl}) - E + u_{fl} \times B] \neq 0.$$

because it implies that the fluid momentum equation is invalid at zero-temperature $var2 = vari = 0$.

The conditions for such a universality requirement are:

$$0 = function(var1, 0); \tag{9.1}$$

$$0 = d_{var2} function. \tag{9.2}$$

Let us see detailed forms of these two conditions. 1. For Eq.(A.1), these two conditions could be expressed by two equations

$$0 = \partial_t[p_{fl} - p_{fl}(u_{fl}, 0)] + u_{fl} \cdot \nabla_r[p_{fl} - p_{fl}(u_{fl}, 0)] + \text{thermal pressure/density}; \tag{A.2.a}$$

$$0 = \partial_t[p_{fl}(u_{fl}, 0)] + u_{fl} \cdot \nabla_r[p_{fl}(u_{fl}, 0)] - [E + u_{fl} \times B]. \tag{A.2.b}$$

2. For Eq.(B.1), we have

$$0 = [\partial_t(f - f_{mono}) + v \cdot \nabla (f - f_{mono}) - [E + v \times B] \cdot \partial_p (f - f_{mono})]$$
$$+ (v - u_{fl}) \cdot \nabla f_{mono} - (v - u_{fl}) \cdot \partial_p f_{mono}; \tag{B.2.a}$$
$$0 = [\partial_t f_{mono} + u_{fl} \cdot \nabla f_{mono} - [E + u_{fl} \times B] \cdot \partial_p f_{mono}]. \tag{B.2.b}$$

3. For Eq.(C.1), we have

$$0 = \left[ \left[ d_t \frac{d_t r_i(t)}{\sqrt{1 - [d_t r_i(t)]^2}} - d_t \frac{u(r_i(t), t)}{\sqrt{1 - [u(r_i(t), t)]^2}} \right] - [d_t r_i(t) - u(r_i(t), t)] \times B(r_i(t), t) \right]; \tag{C.2.a}$$

$$0 = \left[ d_t \frac{u(r_i(t), t)}{\sqrt{1 - [u(r_i(t), t)]^2}} - E(r_i(t), t) - u(r_i(t), t) \times B(r_i(t), t) \right]. \tag{C.2.b}$$

Obviously, Eq.(A.2.b) and Eq.(C.2.b) are same, and Eq.(B.2.b) could also be transformed into them according to standard procedure. Therefore, Eq.(A.2.b) or Eq.(B.2.b) or Eq.(C.2.b)+4 Meqs forms a closed equation set.